\documentstyle[11pt]{article}
\begin{document}

\title{Inhomogeneous String Cosmologies }
\author{\and John D. Barrow and Kerstin E. Kunze \\
Astronomy Centre,\\
University of Sussex,\\
Brighton BN1 9QH\\
U.K.}
\maketitle
\date{}

\begin{abstract}
We present exact inhomogeneous and anisotropic cosmological solutions of
low-energy string theory containing dilaton and axion fields. The spacetime
metric possesses cylindrical symmetry. The solutions describe ever-expanding
universes with an initial curvature singularity and contain known
homogeneous solutions as subcases. The asymptotic form of the solution near
the initial singularity has a spatially-varying Kasner-like form. The
inhomogeneous axion and dilaton fields are found to evolve
quasi-homogeneously on scales larger than the particle horizon. When the
inhomogeneities enter the horizon they oscillate as non-linear waves and the
inhomogeneities attentuate. When the inhomogeneities are small they behave
like small perturbations of homogeneous universes. The manifestation of
duality and the asymptotic behaviour of the solutions are investigated. 

\end{abstract}

\section{Introduction}

The low-energy effective action of the bosonic sector of string theory
provides a gravitation theory containing dilaton and axion fields that
possesses cosmological solutions. These solutions provide models for the
behavior of the universe near the Planck (or string) energy scale [1]. They
allow us to investigate a number of long-standing questions regarding the
occurrence of singularities, the behavior of the general solution of the
theory in the vicinity of a singularity, and the likelihood of our Universe
arising from generic initial data. They also provide a basis for
investigation of higher-order corrections to low-energy cosmological string
theory. Several studies have recently been made of string cosmologies in
order to ascertain the behavior of simple isotropic and anisotropic
universes, investigate the implications of duality, and search for
inflationary solutions [2]-[8]. Since one of the prime reasons for studying
such solutions is to shed light on the behavior of the universe at very high
energies, where our knowledge of its material content, geometrical and
topological properties, or its anisotropies and inhomogeneities, is
necessarily incomplete, it is unwise to make special assumptions about the
form of the cosmological solutions. Indeed, any dimensional reduction
process could be viewed as an extreme form of anisotropic evolution in more
than three dimensions in which three spatial dimensions expand whilst the
rest remain static. A number of studies have focused on obtaining particular
solutions for 3+1 dimensional spacetimes in cases where spatial homogeneity
(and sometimes also isotropy) is assumed for the metric of spacetime, where
the $H$ field is set to zero [4], or where the $H$ field is included by
assuming that it takes a particular form which satisfies its constraints and
its equation of motion [5]. For example, Copeland et al, [2], discussed
Friedmann and Bianchi type I universes, allowing $*H$ to be time dependent
or space dependent, respectively. In a second paper, [3], they discussed
Bianchi I solutions with a homogeneous antisymmetric tensor field. In [6]
(see also [5]) Batakis presented an overview of all possible configurations
of a (spatially) homogeneous $H$-field in diagonal Bianchi models. Whereas,
in ref. [7], we investigated the case for a (spatially) homogeneous tensor
potential $B_{\mu \nu }$ in Bianchi metrics that are not necessarily
diagonal. We also gave a classification of all the degrees of freedom
permitted for the $H$ field in spatially homogeneous universes possessing a
3-parameter group of motions. The only spatially homogeneous universe
excluded from this study is the (closed) $S^2\times S^1$ Kantowski -Sachs
universe. A detailed study of this universe was made by Barrow and Dabrowski
[8].

In this paper, we take one further step upwards in generality and consider a
wide class of inhomogeneous and anisotropic string cosmologies. These
possess cylindrical symmetry and contain homogeneous Bianchi and
Kantowski-Sachs universes as special cases [9]. They allow us to investigate
the propagation of non-linear inhomogeneities in the axion and dilaton
fields. On scales larger than the particle horizon inhomogeneities in the
axion and dilaton fields evolve quasi-homogeneously but when the
inhomogeneities enter the horizon they undergo oscillations and attentuate.
In the limit that the amplitude of the inhomogeneities is small we will
recover the results of perturbation studies of homogeneous string
cosmologies in an appropriate gauge. Besides providing exact descriptions of
the gravitational self-interaction of strongly inhomogeneous axion and
dilaton fields these solutions allow us to investigate the impact of duality
upon the form of the solution in a situation where there exist
characteristic spatial scales.

The string world-sheet action for a closed bosonic string in a background
field including all the massless states of the string as part of the
background is given by, [1], 
\begin{equation}
S=-\frac 1{4\pi \alpha ^{\prime }}\int d^2\sigma \{\sqrt{h}h^{\alpha \beta
}\partial _\alpha X^\mu \partial _\beta X^\nu g_{\mu \nu }(X^\rho )+\epsilon
^{\alpha \beta }\partial _\alpha X^\mu \partial _\beta X^\nu B_{\mu \nu
}(X^\rho )+\alpha ^{\prime }\sqrt{h}\phi (X^\rho )R^{(2)}\}
\end{equation}
where $h^{\alpha \beta }$ is the 2-dimensional worldsheet metric, $R^{(2)}$
the worldsheet Ricci scalar, $\epsilon ^{\alpha \beta }$ the worldsheet
antisymmetric tensor, $B_{\mu \nu }(X^\rho )$ the antisymmetric tensor
field, $g_{\mu \nu }(X^\rho )$ the background spacetime metric (graviton), $%
\phi (X^\rho )$ the dilaton, $\alpha ^{\prime }$ is the inverse string
tension, and the functions $X^\rho (\sigma )$ map the string worldsheet into
the physical D-dimensional spacetime manifold.

For the consistency of string theory it is essential that local scale
invariance holds. Imposing this condition results in equations of motion for
the fields $g_{\mu \nu }$, $B_{\mu \nu }$ and $\phi $ which can be derived
to lowest order in $\alpha ^{\prime }$ from the low-energy effective action
for a vanishing cosmological constant 
\begin{equation}
S=\int d^Dx\sqrt{-g}e^{-\phi }(R+g^{\alpha \beta }\partial _\alpha \phi
\partial _\beta \phi -\frac 1{12}H^{\alpha \beta \gamma }H_{\alpha \beta
\gamma }).
\end{equation}
where the antisymmetric tensor field strength $H_{\alpha \beta \gamma
}=\partial _{[\alpha }B_{\beta \gamma ]}$ is introduced.

In a cosmological context it is generally assumed that by some means all but
four of the 10 or 26 dimensions of spacetime are compactified, leaving an
expanding 3+1-dimensional spacetime ($D=4)$. Since we are interested in
cosmological solutions of the field equations derived from the variation of
this action, we adopt the Einstein frame by making the conformal
transformation 
\begin{equation}
g_{\alpha \beta }\rightarrow e^{-\Phi }g_{\alpha \beta }.
\end{equation}
In this frame the 4-dimensional string field equations and the equations of
motion are given by (Greek indices run $\ 0\leq \alpha ,\beta \leq 3$).

The low-energy effective action in the Einstein frame yields to the
following set of equations ($\kappa ^2\equiv 8\pi G,c\equiv 1$) 
\begin{eqnarray}
R_{\mu \nu }-\frac 12g_{\mu \nu }R &=&\kappa ^2(^{(\Phi )}T_{\mu \nu
}+^{(H)}T_{\mu \nu }) \\
\nabla _\mu (e^{-2\Phi }H^{\mu \nu \lambda }) &=&0 \\
\Box \Phi +\frac 16e^{-2\Phi }H_{\mu \nu \lambda }H^{\mu \nu \lambda } &=&0
\end{eqnarray}

where 
\begin{eqnarray}
^{(\Phi )}T_{\mu \nu } &=&\frac 12(\Phi _{,\mu }\Phi _{,\nu }-\frac 12g_{\mu
\nu }(\partial \Phi )^2) \\
^{(H)}T_{\mu \nu } &=&\frac 1{12}e^{-2\Phi }(3H_{\mu \lambda \kappa }H_\nu
^{\;\;\;\lambda \kappa }-\frac 12g_{\mu \nu }H_{\alpha \beta \gamma
}H^{\alpha \beta \gamma })
\end{eqnarray}

Thus in this frame the problem reduces to the study of inhomogeneous general
relativistic cosmologies containing two gravitationally interacting matter
fields. In the next section we shall manipulate these equations into a
soluble system by introducing a particular inhomogeneous spacetime metric
with cylindrical symmetry with a particular topology. In section 3 we give
exact solutions in cases where one (or both) of the axion and dilaton fields
depends only upon the time variable. In section 4 we consider the case where
both fields depend upon time and space coordinates. In section 5 we
investigate the asymptotic behaviours of these fully inhomogeneous solutions
on scales large and smaller than the horizon. In section 6 we study the
nature of duality in these solutions and the results are discussed in
section 7.

\section{Einstein-Rosen Metric}

Consider the anisotropic and inhomogeneous spacetime metric [10] 
\begin{equation}
ds^2=-e^{2(\chi -\psi )}(d\tau ^2-dr^2)+R(e^{2\psi }dz^2+e^{-2\psi }d\phi ^2)
\end{equation}
where $\chi ,\psi ,R$ are unknown functions of $\tau $ and $r$. Thus $%
\partial /\partial z$ and $\partial /\partial \phi $ are Killing vectors.
Without loss of generality it can be assumed that $0\leq z\leq 1$ and $0\leq
\phi \leq 1$. When $\psi =0$ and $R=e^{2\chi },$ with $R\equiv R(\tau )$ and 
$\chi \equiv \chi (\tau ),$ we recover an isotropic Friedmann universe.
Other homogeneous specialisations of the metric reduce it to one of the
Bianchi type homogeneous universes [9]. Properties of the metric (9) depend
on whether $B_\mu \equiv R_{,\mu }$ is spacelike, timelike or null (Greek
indices run $0\rightarrow 3$). The cases with a globally null or spacelike $%
B_\mu $ correspond to plane or cylindrical gravitational waves, respectively
[2]. Metrics where the sign of $B_\mu B^\mu $ varies throughout the
spacetime describe colliding gravitational waves [3] or cosmologies with
timelike and spacelike singularities [4]. Metrics with a globally timelike $%
B_\mu $ describe cosmological models with spacelike singularities. If the
spacelike hypersurfaces are compact then the allowed spatial topologies [5],
[9] are a 3-torus, $S^1\otimes S^1\otimes S^1,$ for $R=(detg_{ab})^{\frac
12}=\tau $ with $0\leq \tau <\infty $ and $0\leq r<\infty $; a hypertorus, $%
S^1\otimes S^2$, or a 3-sphere, $S^3$, for $R=(detg_{ab})^{\frac 12}=\sin
r\sin \tau $ with $0\leq r\leq \pi $ and $0\leq \tau \leq \pi $. We shall
present solutions for the globally timelike case, $R=\tau .$ These
correspond to ever-expanding cosmological models with an initial curvature
singularity at $\tau =0.$ Note that the behavior of the closed $S^3$ models
approaches that of the 3-torus universes as the singularities are approached
because $\sin \tau \rightarrow \tau $ as $\tau \rightarrow 0$ and $\pi ,$
and so the role played by the duality invariance of these models can be
investigated along with the implications for the 'pre big bang' scenario of
Gasperini et al [4]. The homogeneous models of the $S^1\otimes S^2$ case
will be the Kantowski-Sachs universes studied in ref. [8]. For further
results about the singularity structure and global existence of these
metrics (the strong cosmic censorship hypothesis holds) see the paper by
Chrusciel et al [15].

Rewriting (5) as 
\begin{eqnarray}
d(\star H)-2(d\Phi )\wedge (\star H)=0,
\end{eqnarray}
and using 
\begin{eqnarray}
dH=0,
\end{eqnarray}
we can determine the general form of $H$ that is compatible with the
Einstein-Rosen spacetime geometry. Denoting $x^0=\tau $, $x^1=r$, $x^2=z$,
and $x^3=\phi $ we require 
\begin{eqnarray}
H &=&6A(\tau ,r)dx^0\wedge dx^1\wedge dx^2+6B(\tau ,r)dx^0\wedge dx^1\wedge
dx^3 \\
&&\ \ +6C(\tau ,r)dx^0\wedge dx^2\wedge dx^3+6D(\tau ,r)dx^1\wedge
dx^2\wedge dx^3.  \nonumber
\end{eqnarray}
The quantities $H$ and $\Phi $ can be a functions only of $r$ and $\tau $
here since the energy-momentum tensor is allowed to depend only on these
variables. Hence (11) implies 
\begin{eqnarray}
\partial _1C-\partial _0D=0,
\end{eqnarray}
while $\star H$ is given by 
\begin{equation}
\star H=\frac 16\epsilon _{\mu \nu \lambda \alpha }H^{\mu \nu \lambda
}dx^\alpha \equiv F_\alpha dx^\alpha .
\end{equation}
Since $d\Phi =\partial _0\Phi dx^0+\partial _1\Phi dx^1,$ eqn. (10) reduces
to 
\begin{eqnarray}
\lbrack \partial _0F_1-\partial _1F_0-2(F_1\partial _0\Phi -F_0\partial
_1\Phi )]dx^0\wedge dx^1 &&  \nonumber \\
+[\partial _0F_2-2F_2\partial _0\Phi ]dx^0\wedge dx^2+[\partial
_1F_2-2F_2\partial _1\Phi ]dx^1\wedge dx^2 &&  \nonumber \\
+[\partial _0F_3-2F_3\partial _0\Phi ]dx^0\wedge dx^3+[\partial
_1F_3-2F_3\partial _1\Phi ]dx^1\wedge dx^3 &=&0.
\end{eqnarray}
This implies 
\begin{eqnarray}
\partial _0F_1-2F_1\partial _0\Phi -\partial _1F_0+2F_0\partial _1\Phi &=&0
\\
d(e^{-2\Phi }F_2) &=&0 \\
d(e^{-2\Phi }F_3) &=&0,
\end{eqnarray}
so (17) and (18) yield 
\begin{eqnarray}
F_2 &=&e^{2\Phi }A_2, \\
F_3 &=&e^{2\Phi }A_3,
\end{eqnarray}
where $A_2$ and $A_3$ are constants.

Using the fact that $C(r,\tau
)=g_{00}g_{22}g_{33}H^{023}=g_{00}g_{22}g_{33}\epsilon ^{0231}F_1=-RF_1$
and, similarly that $D(r,\tau )=-RF_0,$ eqn. (13) becomes 
\begin{eqnarray}
\partial _1(RF_1)-\partial _0(RF_0)=0,
\end{eqnarray}
and (16) implies that 
\begin{eqnarray}
\partial _0(e^{-2\Phi }F_1)=\partial _1(e^{-2\Phi }F_0).
\end{eqnarray}

In order to solve the system of differential equations (21)-(22) there are
two obvious choices:

\hspace{0.3cm}(i) $e^{-2\Phi}F_{1}=\partial_{1}b$ and $e^{-2\Phi}F_{0}=%
\partial_{0}b$

\hspace{0.3cm}(ii) $RF_{1}=\partial_{0}h$ and $RF_{0}=\partial_{1}h$

\vspace{0.5 cm}

The latter choice corresponds to taking $B_{23}$ to be the only
non-vanishing component of the antisymmetric tensor potential defined by $%
H=dB$ and depending only on $\tau$ and $r$.

The choice (i) reduces (21) to 
\begin{eqnarray}
\Box b+2\nabla ^\mu b\nabla _\mu \Phi =0,
\end{eqnarray}

while choice (ii) produces another coupled wave equation 
\begin{eqnarray}
\ddot h-h^{\prime \prime }-\frac{\dot R}R\dot h+\frac{R^{\prime }}Rh^{\prime
}-2(\dot \Phi \dot h-\Phi ^{\prime }h^{\prime })=0
\end{eqnarray}

where $\cdot $ $\equiv \frac \partial {\partial \tau }$ and $^{\prime
}\equiv \frac \partial {\partial r}.$

\vspace{0.5cm}

The off-diagonal components of the Einstein tensor $G_{02}$, $G_{03}$, $%
G_{12}$, $G_{13}$ and $G_{23},$ are zero in the spacetime (9). The
corresponding components of $^{(\Phi )}T_{\mu \nu }$ all vanish so we only
need ensure that all the corresponding off-diagonal components of $%
^{(H)}T_{\mu \nu }$ are also zero. Since we have 
\begin{equation}
^{(H)}T_{02}=\frac 12F_2F_0e^{-2\Phi }=\frac 12A_2F_0,
\end{equation}
we must therefore set $A_2=0,$ and hence $H^{013}=B=0.$ For the $(03)$
component we have

\begin{equation}
^{(H)}T_{03}=\frac 12F_3F_0e^{-2\Phi }=\frac 12A_3F_0,
\end{equation}
so we must set $A_3=0,$ and hence $H^{120}=A=0.$ With these choices, the
components $^{(H)}T_{12}$, $^{(H)}T_{13}$ and $^{(H)}T_{23}$ also all
vanish. Therefore, the equations governing the dilaton and antisymmetric
tensor field for the two choices are given by equation (6) together with the
following coupled propagation equations, (27)-(28) or (29)-(30), in the
cases (i) and (ii), respectively:

\hspace{0.3cm}(i) 
\begin{eqnarray}
\frac 1R\ \left( R\Phi ^{\prime }\right) ^{\prime }-\frac 1R\ \left( R\dot
\Phi \right) ^{{\bf \cdot }}-e^{2\Phi }\left[ b^{\prime \;2}-\dot b^2\right]
=0,
\end{eqnarray}

\begin{eqnarray}
\frac 1R\ \left( Rb^{\prime }\right) ^{\prime }-\frac 1R\ \left( R\dot
b\right) ^{{\bf \cdot }}+2\left[ \Phi ^{\prime }b^{\prime }-\dot \Phi \dot
b\right] =0.
\end{eqnarray}

\hspace{0.3cm}(ii) 
\begin{eqnarray}
R^{-1}(R\Phi^{\prime})^{\prime}-R^{-1}(R\dot{\Phi})^{.}+R^{-2}e^{-2\Phi}[
h'^{\;2}-\dot{h}^{2}]=0, \\
\ddot h-h^{\prime \prime }-\frac{\dot R}R\dot h+\frac{R^{\prime }}Rh^{\prime
} =-2(\Phi ^{\prime }h^{\prime }-\dot \Phi \dot h).
\end{eqnarray}

Since both choices involve the same number of independent functions they are
equivalent; here, choice (i) is taken.

The energy-momentum tensor in (4) reads 
\begin{equation}
{\kappa }^2\;^{(\Phi )}T_\mu ^{\;\;\lambda }=\frac 12(g^{\nu \lambda
}\partial _\mu \Phi \partial _\nu \Phi -\frac 12\delta _\mu ^\lambda
(\partial \Phi )^2),
\end{equation}
\begin{equation}
{\kappa }^2\;^{(H)}T_\mu ^{\;\;\lambda }=\frac 12e^{2\Phi }(g^{\nu \lambda
}\partial _\mu b\partial _\nu b-\frac 12\delta _\mu ^\lambda (\partial b)^2),
\end{equation}
so the nonvanishing components of the energy-momentum tensor are 
\begin{eqnarray}
\kappa ^2T_0^{\;\;0} &=&-\frac 14e^{-2(\chi -\psi )}[\dot \Phi ^2+\Phi
^{\prime }{}^{\;2}+(\dot b^2+b^{\prime }{}^{\;2})e^{2\Phi }]=-\kappa
^2T_1^{\;1},  \nonumber \\
\kappa ^2T_0^{\;\;1} &=&\frac 12e^{-2(\chi -\psi )}[\dot \Phi \Phi ^{\prime
}+\dot bb^{\prime }e^{2\Phi }]=-\kappa ^2T_1^{\;0},  \nonumber \\
\kappa ^2T_2^{\;\;2} &=&\frac 14e^{-2(\chi -\psi )}[\dot \Phi ^2-\Phi
^{\prime }{}^{\;2}+(\dot b^2-b^{\prime }{}^{\;2})e^{2\Phi }]=\kappa
^2T_3^{\;3},  \nonumber \\
&&
\end{eqnarray}

The energy-momentum tensor can be interpreted as describing two stiff
perfect fluids where the energy density for the dilaton fluid is found to be 
\[
p_\Phi =\rho _\Phi =\frac 14e^{-2(\chi -\psi )}[\dot \Phi ^2-\Phi ^{\prime
}{}^{\;2}] 
\]
and its 4-velocity is given by 
\[
u^\alpha =e^{-(\chi -\psi )}[\dot \Phi ^2-\Phi ^{\prime }{}^{\;2}]^{-\frac
12}(-\dot \Phi ,\Phi ^{\prime },0,0) 
\]
and for the axion fluid we have 
\[
p_H=\rho _H=\frac 14e^{-2(\chi -\psi )}[\dot b^2-b^{\prime
}{}^{\;2}]e^{2\Phi } 
\]
and its 4-velocity is 
\[
v^\alpha =e^{-(\chi -\psi )}[\dot b^2-b^{\prime }{}^{\;2}]^{-\frac 12}(-\dot
b,b^{\prime },0,0). 
\]

Furthermore, $T_0^{\;\;0}+T_1^{\;\;1}=0$ and $T_2^{\;\;2}-T_3^{\;\;3}=0$ as
the cylindrical symmetry of the metric demands, and Einstein's equations for 
$R(\tau ,r)$ and $\psi (\tau ,r)$ are given by the linear wave equations
[11] 
\begin{eqnarray}
\ddot R-R^{\prime \prime } &=&0, \\
\ddot \psi +\frac{\dot R}R\dot \psi -\psi ^{\prime \prime }-\frac{R^{\prime }%
}R\psi ^{\prime } &=&0.
\end{eqnarray}

The remaining metric function, $\chi (\tau ,r),$ is determined by the two
Einstein constraint equations

\begin{eqnarray}
\chi ^{\prime } &=&\psi ^{\prime }-\frac 14\frac{R^{\prime }}R-(\dot
R^2-R^{\prime }{}^{\;2})^{-1}[RR^{\prime }(\dot \psi ^2+\psi ^{\prime
}{}^{\;2})+R^{\prime }R^{\prime \prime }-2\dot RR\dot \psi \psi ^{\prime
}-\dot R\dot R^{\prime } \\
&&-\kappa^{2} Re^{2(\chi -\psi )}(T^{\;\;0}_0R^{\prime }+T_{0}^{\;\;1}\dot
R)],  \nonumber \\
\dot \chi &=&\dot \psi -\frac 14\frac{\dot R}R-(\dot R^2-R^{\prime
}{}^{\;2})^{-1}[2RR^{\prime }\dot \psi \psi ^{\prime }-R\dot R(\dot \psi
^2+\psi ^{\prime }{}^{\;2})-\dot RR^{\prime \prime }+R^{\prime }\dot
R^{\prime } \\
&&+\kappa^{2} e^{2(\chi -\psi )}R(T^{\;\;0}_0\dot R+T_{0}^{\;\;1}R^{\prime
})].  \nonumber
\end{eqnarray}

Since cosmological solutions are of interest to us here, we consider only
the timelike solution of (34). Using the general coordinate invariances $%
(\tau \pm r\rightarrow f(\tau \pm r))$ of the metric this may be taken
without loss of generality to be

\begin{equation}
R=R(\tau )=\tau .
\end{equation}
Then, equation (35) reduces to 
\begin{eqnarray}
\ddot \psi +\frac 1\tau \dot \psi -\psi ^{\prime \prime }=0
\end{eqnarray}
which is solved by 
\begin{eqnarray}
\psi (\tau ,r)=^0\psi _0+^0\psi _1\ln \tau +\sum_{n=1}^\infty \cos
[n(r-r_n)][^A\Psi _nJ_0(n\tau )+^B\Psi _nN_0(n\tau )]
\end{eqnarray}
where $^0\psi _i$, $^A\Psi _n$, $^B\Psi _n$, $r_n$ are constants and $J_0(x)$
and $N_0(x)$ denote the 0$^{th}$ order Bessel and Neumann functions,
respectively.

Equations (27)-(28) read 
\begin{eqnarray}
\ \Phi ^{\prime \prime }-\frac{\dot \Phi }\tau \ -\ddot \Phi -e^{2\Phi
}\left[ b^{\prime \;2}-\dot b^2\right] &=&0 \\
b^{\prime \prime }-\frac{\dot b}\tau -\ddot b+2\left[ \Phi ^{\prime
}b^{\prime }-\dot \Phi \dot b\right] &=&0
\end{eqnarray}

In the next section several solutions will be found.

\section{Solutions of varying generality}

Before explicit solutions are given something general we make some remarks
about the procedure for solving the system of partial differential equations
for the metric function $\chi (\tau ,r)$.

For $R(\tau )=\tau ,$ eqns. (36) \& (37) reduce to 
\begin{eqnarray}
\chi ^{\prime } &=&\psi ^{\prime }+2\tau \dot \psi \psi ^{\prime }+\kappa
^2\tau e^{2(\chi -\psi )}T_0^{\;\;1} \\
\dot \chi &=&\dot \psi -\frac 1{4\tau }+\tau (\dot \psi ^2+\psi ^{\prime
}{}^{\;2})-\kappa ^2\tau e^{2(\chi -\psi )}T_0^{\;\;0}
\end{eqnarray}

Generally speaking, the most difficult step is to find the integral for the
part coupled to $\psi (\tau ,r)$. However, this problem was solved by
Charach [16]. Define a function $G(\psi ;\tau ,r)$ by 
\begin{eqnarray}
G^{\prime } &=&\tau \dot \psi \psi ^{\prime } \\
\dot G &=&\frac 12\tau (\dot \psi ^2+\psi ^{\prime }{}^{\;2})
\end{eqnarray}

Note that $\psi $ satisfying 
\[
\ddot \psi +\frac 1\tau \dot \psi -\psi ^{\prime \prime }=0 
\]
is kept as a functional dependence in $G$. The explicit dependence on $\tau $
and $r$ might sometimes be suppressed and we write $G(\psi )$ as

\begin{eqnarray}
G(\psi;\tau, r) &=&^{0}\psi_{0}+\frac 12\ (^{0}\psi_{1})^2\ln\tau  \nonumber
\\
&&\ \ \ \ \ \ \ +^{0}\psi_{1}\sum_{n=1}^\infty \cos [n(r-r_n)][
^{A}\Psi_{n}J_0(n\tau )+ ^{B}\Psi_{n} N_0(n\tau )]  \nonumber \\
&&\ \ \ \ \ \ \ +\frac 14\tau ^2\sum_{n=1}^\infty n^2([
^{A}\Psi_{n}J_0(n\tau )+ ^{B}\Psi_{n}N_0(n\tau )]^2+[ ^{A}\Psi_{n}J_1(n\tau
)+ ^{B}\Psi_{n}N_1(n\tau )]^2)  \nonumber \\
&&\ \ \ \ \ \ \ -\frac 12\tau \sum_{n=1}^\infty n\cos
^2[n(r-r_n)]\{(^{A}\Psi_{n})^2J_0(n\tau )J_1(n\tau )  \nonumber \\
&&\ \ \ \ +^{A}\Psi_{n}^{B}\Psi_{n}[N_0(n\tau )J_1(n\tau )+J_0(n\tau
)N_1(n\tau )]+ (^{B}\Psi_{n})^2 N_0(n\tau )N_1(n\tau )\}  \nonumber \\
&&\ \ \ \ +\frac 12\tau \sum_{n=1}^\infty \sum_{m=1,n\neq m}^\infty \frac{nm%
}{n^2-m^2}\{\sin [n(r-r_n)]\sin [m(r-r_m)][nU_{nm}^{(0)}(\tau
)-mU_{nm}^{(1)}(\tau )]  \nonumber \\
&&\ \ \ +\cos [n(r-r_n)]\cos [m(r-r_m)][mU_{nm}^{(0)}(\tau
)-nU_{nm}^{(1)}(\tau )]\}  \nonumber \\
&&
\end{eqnarray}
where 
\begin{eqnarray}
&&  \nonumber \\
U_{nm}^{(0)}(\tau ) &\equiv ^{A}\Psi_{n} ^{A}\Psi_{m}J_1(n\tau )J_0(n\tau )+
^{B}\Psi_{n} ^{B}\Psi_{m}N_0(m\tau )N_1(n\tau )+2^{A}\Psi_{n}
^{B}\Psi_{m}J_1(n\tau )N_0(m\tau ),  \nonumber
\end{eqnarray}
\begin{eqnarray}
U_{nm}^{(1)}(\tau )\equiv ^{A}\Psi_{n} ^{A}\Psi_{m}J_0(n\tau )J_0(m\tau )
+^{B}\Psi_{n} ^{B}\Psi_{m}N_0(n\tau )N_1(m\tau )+2 ^{A}\Psi_{n}
^{B}\Psi_{m}J_0(n\tau )N_1(m\tau ).  \nonumber
\end{eqnarray}

We now consider classes of solutions in which one (or both) of the $\Phi $
and $b$ fields depend on only one of the coordinates $r$ and $t.$

\subsection{Solutions homogeneous in $\tau$ : $\Phi =\Phi (\tau )$, $%
b=b(\tau )$}

The well-known solution to (41) \& (42) [6] in this subcase is 
\begin{eqnarray}
e^\Phi &=&\cosh (N\zeta )+\sqrt{1-\frac{B^2}{N^2}}\sinh (N\zeta ) \\
b(\zeta ) &=&\frac NB\frac{\sinh (N\zeta )+\sqrt{1-\frac{B^2}{N^2}}\cosh
(N\zeta )}{\cosh (N\zeta )+\sqrt{1-\frac{B^2}{N^2}}\sinh (N\zeta )}
\end{eqnarray}
where $N$ and B are constants and $d\tau =\tau d\zeta$.

Using this in the expression for the components of the energy-momentum
tensor gives an expression for $\chi (\tau ,r)$ 
\begin{eqnarray}
\chi (\tau ,r)=\psi (\tau ,r)+2G(\psi ;\tau ,r)+\frac{N^2-1}4\ln \tau +M,
\end{eqnarray}

where $M$ is a constant. Hence, the metric function $\exp [\chi -\psi ]$ is
given by

\begin{eqnarray}
\exp[\chi-\psi]=e^{2G(\psi)}e^{M}\tau^{\frac{N^{2}-1}{4}}
\end{eqnarray}

\subsection{Solutions homogeneous in $r$ : $\Phi =\Phi (r)$, $b=b(r)$}

The solution in this subcase is given by 
\begin{eqnarray}
e^\Phi &=&\cosh (Nr)+\sqrt{1-\frac{B^2}{N^2}}\sinh (Nr) \\
b(r) &=&\frac NB\frac{\sinh (Nr)+\sqrt{1-\frac{B^2}{N^2}}\cosh (Nr)}{\cosh
(Nr)+\sqrt{1-\frac{B^2}{N^2}}\sinh (Nr)}
\end{eqnarray}
where $N$, $B$ constants. From these expressions, $\chi (\tau ,r)$ is found
to be

\begin{eqnarray}
\chi (\tau ,r)=\psi (\tau ,r)+2G(\psi ;\tau ,r)-\frac 14\ln \tau +\frac{N^2}%
8\tau ^2+M,
\end{eqnarray}
which gives the remaining metric component,

\begin{eqnarray}
\exp [\chi -\psi ]=e^{2G(\psi )}e^{2M}\tau ^{-\frac 14}e^{\frac{N^2}8\tau
^2}.
\end{eqnarray}

\subsection{Solutions with an oscillatory axion : $\Phi =\Phi (\tau )$, $%
b=b(\tau,r)$}

If we rewrite (42) as 
\begin{equation}
\frac{\partial ^2b}{\partial r^2}-\frac{\partial ^2b}{\partial \tau ^2}-%
\frac{\partial b}{\partial \tau }(2\frac{\partial \Phi }{\partial \tau }%
+\frac 1\tau )+2\frac{\partial \Phi }{\partial r}\frac{\partial b}{\partial r%
}=0,
\end{equation}
and take a solution $2\Phi (\tau )=-\ln (\tau /\tau _0),$ then the axion
field $b(r,\tau )$ also satisfies the wave equation, 
\begin{equation}
\frac{\partial ^2b}{\partial r^2}-\frac{\partial ^2b}{\partial \tau ^2}=0,
\end{equation}
which has the general solution 
\begin{equation}
b(r,\tau )=\alpha b_1(r+\tau )+\beta b_2(r-\tau )
\end{equation}
where $\alpha $, $\beta $ are constants and $b_i$ are arbitrary functions of
their arguments. Eqn. (41) is satisfied if 
\begin{equation}
\left( \frac{\partial b}{\partial r}\right) ^2-\left( \frac{\partial b}{%
\partial \tau }\right) ^2=0.
\end{equation}
This implies 
\begin{equation}
\left( \frac{\partial b}{\partial r}+\frac{\partial b}{\partial \tau }%
\right) \left( \frac{\partial b}{\partial r}-\frac{\partial b}{\partial \tau 
}\right) =0
\end{equation}
so that either $\alpha $ or $\beta $ must vanish. Thus we obtain the
solution 
\begin{eqnarray}
\Phi (\tau ) &=&-\frac 12\ln \frac \tau {\tau _0}, \\
b(\tau ,r) &=&\Theta (u)b_1(r+\tau )+(1-\Theta (u))b_2(r-\tau ).
\end{eqnarray}
where $\Theta (u)$ is the step function ($\Theta (u)=0$ for $u\leq 0;\Theta
(u)=1$ for $u>0$) and $u$ an arbitrary real parameter.

It is interesting to have
a solution with a homogeneous dilaton and an {\sl in}homogeneous axion.
Note, that in this case the axion behaves quite differently from the dilaton.

The function $\chi (\tau ,r)$ is determined by

\begin{eqnarray}
\chi ^{\prime } &=&\psi ^{\prime }+2\tau \dot \psi \psi ^{\prime }+\frac{%
\tau _0}2\dot bb^{\prime } \\
\dot \chi &=&\dot \psi -\frac 3{16\tau }+\tau (\dot \psi ^2+\psi ^{\prime
}{}^{\;2})+\frac{\tau _0}4(\dot b^2+b^{\prime }{}^{\;2}).
\end{eqnarray}

To solve this system of equations we define a new function $B(\tau ,r)$
satisfying 
\begin{eqnarray}
B^{\prime } &=&\dot bb^{\prime } \\
\dot B &=&\frac 12(\dot b^2+b^{\prime }{}^{\;2})
\end{eqnarray}
Changing to new variables, 
\[
X=r+\tau \;,\;\;\;\;\;\;\;Y=r-\tau , 
\]
we find 
\[
\frac{\partial B}{\partial X}=\left( \frac{\partial b}{\partial X}\right)
^2\;\;,\;\;\frac{\partial B}{\partial Y}=-\left( \frac{\partial b}{\partial Y%
}\right) ^2. 
\]
This implies 
\begin{eqnarray}
\frac{\partial ^2B}{\partial X\partial Y}=0
\end{eqnarray}
which is generally solved by 
\begin{eqnarray}
B(X,Y)=B_1(X)+B_2(Y)
\end{eqnarray}
with $B_i$ arbitrary functions of their arguments. Using the general
solution for $b$ in terms of $X$ and $Y$, $B(X,Y)$ is given by

\begin{eqnarray}
B(X,Y)=\Theta (u)\int dX\left( \frac{db_1}{dX}\right) ^2-(1-\Theta (u))\int
dY\left( \frac{db_2}{dY}\right) ^2
\end{eqnarray}
Finally, an expression for $\chi (\tau ,r)$ is obtained, 
\begin{eqnarray}
\chi (\tau ,r)=\psi (\tau ,r)+2G(\psi )+\frac{\tau _0}2B(\tau ,r)-\frac
3{16}\ln \tau +M,
\end{eqnarray}
which results in

\begin{eqnarray}
\exp [\chi -\psi ]=e^{2G(\psi )}e^M\tau ^{-\frac 3{16}}e^{\frac{\tau _0}%
2B(\tau ,r)}.
\end{eqnarray}

\subsection{Solutions with $\Phi =\Phi (\tau, r )$, $b=b(\tau )$}

If we take $b(\tau )=A\tau ^2/2$, $A$ constant, and $\Phi (\tau ,r)=-\ln
\tau +S(r)$, then eqn. (41) requires $S(r)$ to satisfy 
\[
\frac{d^2S}{dr^2}+A^2e^{2S(r)}=0. 
\]
Hence, 
\begin{eqnarray}
e^{-\Phi (r,\tau )} &=&\tau [\cosh (Nr)+\sqrt{1-\frac{A^2}{N^2}}\sinh (Nr)],
\\
b(\tau ) &=&\frac A2\tau ^2.
\end{eqnarray}
is a solution of (41)-(42).

Calculating the appropriate components of the energy-momentum tensor yields

\begin{eqnarray}
\chi (\tau ,r)=\psi (\tau ,r)+2G(\psi ;\tau ,r)+\frac 12\ln [\cosh (Nr)+%
\sqrt{1-\frac{A^2}{N^2}}\sinh (Nr)]+\frac{N^2}8\tau ^2+M,
\end{eqnarray}
hence

\begin{eqnarray}
\exp [\chi -\psi ]=e^{2G(\psi )}e^M[\cosh (Nr)+\sqrt{1-\frac{A^2}{N^2}}\sinh
(Nr)]^{\frac 12}e^{\frac{N^2}8\tau ^2}.
\end{eqnarray}

\subsection{Discussion}

Apart from case (3.3) the solutions presented so far describe
non-oscillatory axion-dilaton systems on an oscillatory cosmological
background. In case (3.3) the axion field is allowed to oscillate which
couples the dilatonic and gravitational waves. However, because of condition
(59), only travelling wave solutions in $b(\tau ,r)$ are described in this
case.

\section{Charach Solutions}

The system of equations (41)-(42) is very similar to equations determining
the components of the electromagnetic potential in the electromagnetic Gowdy
universe [16] [17]. It was stated in [16] (and references therein) that the
geometric requirements of the Einstein-Rosen spacetimes allow four
independent components of the 6 possible components of the Maxwell tensor
which can be derived from two non-vanishing components of the
electromagnetic potential. In section 2, we found that only two of the four
possible components of the antisymmetric tensor field strength can be
non-vanishing, which can then be accordingly derived from the potential like
function $b(\tau ,r)$ or $h(\tau ,r$). In the latter case there is a direct
connection to the antisymmetric tensor field potential $B_{\mu \nu }$ where $%
H=dB$. In order to obtain an exact solution of (41)-(42) where $\Phi $ and $%
b $ are dependent on $r$ and $\tau $ we employ a procedure introduced by
Charach [16].

Assume that 
\begin{eqnarray}
\Phi (r,\tau )=-\frac 12\ln v[b(r,\tau )]
\end{eqnarray}
where $v(b)$ is a function yet to be determined. Since 
\[
\dot \Phi =-\frac 12\dot b\frac{d\ln v}{db},\;\;\;\;\;\;\;\;\;\Phi ^{\prime
}=-\frac 12b^{\prime }\frac{d\ln v}{db}, 
\]
\[
\ddot \Phi =-\frac 12[\ddot b\frac{d\ln v}{db}+\dot b^2\frac{d^2\ln v}{db^2}%
],\;\;\;\;\;\;\;\;\Phi ^{\prime \prime }=-\frac 12[b^{\prime \prime }\frac{%
d\ln v}{db}+b^{\prime }{}^{\;2}\frac{d^2\ln v}{db^2}], 
\]
we can use (42) to transform eqn. (41) into 
\begin{eqnarray}
\left( \frac{d^2v}{db^2}+2\right) \left( b^{\prime }{}^{\;2}-\dot b^2\right)
=0,
\end{eqnarray}
while eqn. (42) becomes 
\begin{eqnarray}
b^{\prime \prime }-\ddot b-\frac 1\tau \dot b=[b^{\prime }{}^{\;2}-\dot b^2]%
\frac{d\ln v}{db}.
\end{eqnarray}
Assuming $b^{\prime }{}^{\;2}-\dot b^2\neq 0$, eqn. (77) implies 
\begin{eqnarray}
v(b)=-b^2+c_1b+c_2,
\end{eqnarray}
where the logarithm in (76) requires that the constants $c_1$, $c_2$ satisfy
the inequality 
\begin{equation}
c_1^2+4c_2>0.
\end{equation}
Equation (78) becomes 
\begin{eqnarray}
b^{\prime \prime }-\ddot b-\frac 1\tau \dot b=\frac{c_1-2b}{c_2+c_1b-b^2}%
\left( b^{\prime }{}^{\;2}-\dot b^2\right) .
\end{eqnarray}
If we make the substitution 
\begin{eqnarray}
b=b_0+M\tanh (M\omega )
\end{eqnarray}
eqn. (81) becomes

\begin{equation}
\frac{M^2}{\cosh ^2(M\omega )}[\omega ^{\prime \prime }-\ddot \omega -\frac
1\tau \dot \omega ]+2\frac{M^3\sinh (M\omega )}{\cosh ^3(M\omega )}[\dot
\omega ^2-\omega ^{\prime }{}^{\;2}]=2\frac{M^3\sinh (M\omega )}{\cosh
^3(M\omega )}[\dot \omega ^2-\omega ^{\prime }{}^{\;2}],
\end{equation}
where $b_0=\frac 12c_1$ and $M^2=c_2+\frac 14c_1^2>0.$ Hence, $\omega
(r,\tau )$ satisfies the linear wave equation,

\begin{eqnarray}
\omega ^{\prime \prime }-\ddot \omega -\frac 1\tau \dot \omega =0.
\end{eqnarray}

The wave-packet solution to (84) is given by 
\begin{eqnarray}
\omega (\tau ,r)=^0\omega _0+^0\omega _1\ln \tau +\sum_{n=1}^\infty \cos
[n(r-r_n)][^A\Omega _nJ_0(n\tau )+^B\Omega _nN_0(n\tau )]
\end{eqnarray}
where $^0\omega _i$, $^A\Omega _n,$ $^B\Omega _n$, and $r_n$ are constants.

In summary, equations (41) and (42) admit the following inhomogeneous
solution: 
\begin{eqnarray}
\Phi (\tau ,r) &=&\ln \frac{\cosh (M\omega )}M, \\
b(\tau ,r) &=&\frac 12c_1+M\tanh (M\omega ).
\end{eqnarray}
Rewriting the components $T_0^{\;\;0}$ and $T_1^{\;\;0}$ of the
energy-momentum tensor in terms of $\omega (\tau ,r)$ gives 
\begin{equation}
\kappa ^2e^{2(\chi -\psi )}T_0^{\;\;0}=-\frac{M^2}4(\dot \omega ^2+\omega
^{\prime }{}^{\;2}),
\end{equation}
\begin{equation}
\kappa ^2e^{2(\chi -\psi )}T_0^{\;\;1}=\frac{M^2}2\dot \omega \omega
^{\prime },
\end{equation}
and equations (43) and (44), which determine $\chi (\tau ,r),$ reduce to 
\begin{eqnarray}
\chi ^{\prime } &=&\psi ^{\prime }+2\tau \dot \psi \psi ^{\prime }+\frac{M^2}%
2\tau \dot \omega \omega ^{\prime } \\
\dot \chi &=&\dot \psi -\frac 1{4\tau }+\tau (\dot \psi ^2+\psi ^{\prime
}{}^{\;2})+\frac{M^2}4\tau (\dot \omega ^2+\omega ^{\prime }{}^{\;2})
\end{eqnarray}

Using the function $G(f;\tau ,r)$ where $\ddot f+\tau ^{-1}\dot f-f^{\prime
\prime }=0$ and $G(f;\tau ,r)$ is given by eqn. (47) and eqns. (90)-(91)
lead to 
\begin{eqnarray}
d\chi =d\psi -\frac 14d\ln \tau +2dG(\psi ;\tau ,r)+\frac{M^2}2dG(\omega
;\tau ,r)
\end{eqnarray}
which yields 
\begin{eqnarray}
\chi (\tau ,r)=\psi (\tau ,r)-\frac 14\ln \tau +2G(\psi ;\tau ,r)+\frac{M^2}%
2G(\omega ;\tau ,r)+L
\end{eqnarray}
where $L$ is some constant.

So that the metric function $\exp[\chi-\psi]$ is given by 
\begin{eqnarray}
\exp[\chi-\psi]=e^{L}\tau^{-\frac{1}{4}}e^{2G(\psi)+\frac{M^{2}}{2}G(\omega)}
\end{eqnarray}

\section{Asymptotic behavior}

The existence of inhomogeneity in the solutions found in section 4
introduces characteristic length scales and the gravitational
self-interaction of the dilatonic and axionic waves will differ over scales
according as they are causally coherent or not. The horizon distance in the $%
r$ direction is defined by $ds^2\mid _{z,\phi }=0;$ hence 
\begin{equation}
\triangle r=\int_0^\tau d\tau =\tau .
\end{equation}
Therefore, the combination $n\tau $ in the solutions above can be
interpreted as the ratio of the radial horizon distance to the coordinate
wavelength, $\lambda ,$ since $n\propto 1/\lambda (n)$. There are two
limiting cases to be considered: the case $n\tau \ll 1,$ when the comoving
wavelength is much larger than the radial horizon scale, and the case $n\tau
\gg 1,$ when the wavelength of the inhomogeneities is well within the
horizon scale. We consider these two cases separately.

The Charach solutions discussed in the last section are the most general
ones of those given. Apart from solutions of section (3.3) the limiting
properties of the other solutions are included in those of the Charach-type
solutions. Therefore in this section only the asymptotes of these solutions,
of section 4, are discussed. Explicit formulae for the functions involved
are given in the appendix along with some useful definitions. It is
convenient to define metric functions 
\begin{eqnarray}
A_1(\tau ,r) &\equiv &\exp [\chi (\tau ,r)-\psi (\tau ,r)] \\
A_2(\tau ,r) &\equiv &\tau ^{\frac 12}e^{\psi (\tau ,r)} \\
A_3(\tau ,r) &\equiv &\tau ^{\frac 12}e^{-\psi (\tau ,r)}
\end{eqnarray}

\subsection{The limit $n\tau\ll 1$}

In this case 
\begin{eqnarray}
A_1(\tau ,r) &\sim &e^L\tau ^{-\frac 14+2\gamma _2(\psi ;r)+\frac{M^2}%
2\gamma _2(\omega ;r)}\exp [2\gamma _1(\psi ;r)+\frac{M^2}2\gamma _1(\omega
;r)] \\
A_2(\tau ,r) &\sim &e^{\alpha _1(\psi ;r)}\tau ^{\frac 12+\alpha _2(\psi ;r)}
\\
A_3(\tau ,r) &\sim &e^{-\alpha _1(\psi ;r)}\tau ^{\frac 12-\alpha _2(\psi
;r)}
\end{eqnarray}

This limit corresponds to the case where the comoving wavelength is much
larger than the (radial) horizon size or in other words the universe
consists of causally disconnected regions. In this case one would not expect
to have any oscillatory behavior in $n\tau $.

Concentrating on the homogeneous limit for $\tau$ approaching zero allows to
discuss cosmological solutions near the singularity. The metric functions
are found to approach

\begin{eqnarray}
A_1(\tau ) &\sim &\tau ^{-\frac 14+(^0\psi _1)^2+\frac{M^2}4(^0\omega _1)^2}
\\
A_2(\tau ) &\sim &\tau ^{\frac 12+^0\psi _1} \\
A_3(\tau ) &\sim &\tau ^{\frac 12-^0\psi _1}.
\end{eqnarray}

Changing to proper time using, in the homogeneous limit, the relation 
\[
t=\int d\tau A_1(\tau ), 
\]
$\tau (t)$ is found to be 
\begin{eqnarray}
\tau \propto t^{\frac 1{\frac 34+(^0\psi _1)^2+\frac{M^2}4(^0\omega _1)^2}}.
\end{eqnarray}

Defining the Kasner-exponents $p_i$, $i$=1, 2, 3, by 
\[
g_{\mu \nu }\sim diag(-1,t^{2p_1},t^{2p_2},t^{2p_3}) 
\]
they are found to be slowly spatially varying: 
\begin{eqnarray}
p_1 &\equiv &\frac{-\frac 14+(^0\psi _1)^2+\frac{M^2}4(^0\omega _1)^2}{\frac
34+(^0\psi _1)^2+\frac{M^2}4(^0\omega _1)^2}, \\
p_2 &\equiv &\frac{\frac 12+\,^0\psi _1}{\frac 34+(^0\psi _1)^2+\frac{M^2}%
4(^0\omega _1)^2}, \\
p_3 &\equiv &\frac{\frac 12-\,^0\psi _1}{\frac 34+(^0\psi _1)^2+\frac{M^2}%
4(^0\omega _1)^2}.
\end{eqnarray}
They satisfy the algebraic constraints,

\begin{eqnarray}
\sum_{i=1}^3p_i &=&1, \\
\sum_{i=1}^3p_i^2 &=&1-\frac{M^2}2\frac{(^0\omega _1)^2}{[\frac 34+(^0\psi
_1)^2+\frac{M^2}4(^0\omega _1)^2]^2}.
\end{eqnarray}

The fact that $\sum_{i=1}^3p_i^2\leq 1$, where the equality holds in the
vacuum case ($M=0$), shows immediately that there are isotropic solutions.
This feature is present in the matter-filled Gowdy solutions [16] [17] and
in the spatially homogeneous Kasner universes containing stiff fluid.

The axion-dilaton system is independent of the gravitational background in
the sense that its determining equations (cf (41) and (42)) do not involve
any of the metric functions apart from $R(\tau ,r)$. However, due to the
general structure of the equations the solutions for $\Phi $ and $b$ are
very similar to those of the metric functions. As $\tau \rightarrow 0$, the
dilaton and axion fields approach

\begin{eqnarray}
\Phi (\tau ,r) &\sim &\ln [e^{M\alpha _1(\omega ;r)}\tau ^{M\alpha _2(\omega
;r)}+e^{-M\alpha _1(\omega ;r)}\tau ^{-M\alpha _2(\omega ;r)}]-\ln 2M, \\
b(\tau ,r) &\sim &\frac 12c_1+M\frac{e^{2M\alpha _1(\omega ;r)}\tau
^{2M\alpha _2(\omega ;r)}-1}{e^{2M\alpha _1(\omega ;r)}\tau ^{2M\alpha
_2(\omega ;r)}+1}.
\end{eqnarray}

We note that the early-time behaviour of these solutions falls under the
category of 'velocity-dominated' solutions used in studies of general
relativistic cosmologies [15]. As the singularity is approached the spatial
gradients become negligible with respect to the time derivatives,
3-curvature anisotropies are ignored, and velocities are assumed to less
than the speed of light. This approximation does not encompass the most
general known behaviour in general relativity, with the metric undergoing
chaotic oscillations on approach to the singularity [24]. Chaos in string
cosmologies will the subject of a separate study [25].

\subsection{The limit $n\tau\gg 1$}

In this case the comoving wavelength is smaller than the (radial) horizon
size allowing interaction between different modes and hence an oscillatory
behavior of the metric components. From the limits of $\psi (\tau ,r)$ and $%
G(\psi ;\tau ,r)$ given in the appendix it can be seen that $\psi $ displays
an oscillatory behavior whilst the oscillations in $G$ are damped out.

\begin{eqnarray}
A_1(\tau ,r) &\sim &e^L\tau ^{-\frac 14+2\gamma _4(\psi )+\frac{M^2}2\gamma
_4(\omega )}\exp [2\gamma _3(\psi )+\frac{M^2}2\gamma _3(\omega )]  \nonumber
\\
&&\times \exp [(2\gamma _5(\psi )+\frac{M^2}2\gamma _5(\omega ))\tau ] \\
A_2(\tau ,r) &\sim &\tau ^{\frac 12+\beta _2(\psi )}e^{\beta _1(\psi )}\exp
[\tau ^{-\frac 12}h(\psi ;\tau ,r)] \\
A_3(\tau ,r) &\sim &\tau ^{\frac 12-\beta _2(\psi )}e^{-\beta _1(\psi )}\exp
[-\tau ^{-\frac 12}h(\psi ;\tau ,r)]
\end{eqnarray}

As can be easily seen from the definition of $h(\psi ;\tau ,r)$ given in the
appendix it satisfies the wave equation (in Minkowski space), 
\begin{eqnarray}
\ddot h-h^{\prime \prime }=0.
\end{eqnarray}

The exponential in $A_1$ ensures that the homogeneous limit is approached at
large $\tau ,$ and is an anisotropic universe which can be at most
axisymmetric ($\beta _2(\psi ;t,r)=0$). Since $g_{\mu \nu }\sim
diag(-A_1^{\;2},A_1^{\;2},A_2^{\;2},A_3^{\;2})$ and, for large values of $%
\tau ,$ we have 
\[
\exp [2\tau ^{-\frac 12}h(\psi ;\tau ,r)]\sim 1+2\tau ^{-\frac 12}h[\psi
;\tau ,r], 
\]
and so $g_{\mu \nu }$ can be written as the sum of a background part $\eta
_{\mu \nu }$ and a ``wave'' part $h_{\mu \nu }$ 
\[
g_{\mu \nu }=\eta _{\mu \nu }+h_{\mu \nu }, 
\]
which are found to be

\begin{eqnarray}
\eta _{\mu \nu } &\equiv &diag(-A_1^{\;2},A_1^{\;2},\tau ^{1+2\beta _2(\psi
)}e^{2\beta _1(\psi )},\tau ^{1-2\beta _2(\psi )}e^{-2\beta _1(\psi )}) \\
h_{\mu \nu } &\equiv &diag(0,0,2\tau ^{\frac 12+2\beta _2(\psi )}e^{2\beta
_1(\psi )}h(\psi ;\tau ,r),-2\tau ^{\frac 12-2\beta _2(\psi )}e^{-2\beta
_1(\psi ;t,r)}h(\psi ;\tau ,r))
\end{eqnarray}

The dilaton-axion system displays an oscillatory behavior as well, although
as emphasized before, there is no interaction between gravitational and
axion-dilaton waves.

The asymptotes are given by

\begin{eqnarray}
\Phi (\tau ,r) &\sim &\ln \left[ e^{M\beta _1(\omega )}\tau ^{M\beta
_2(\omega )}+e^{-M\beta _1(\omega )}\tau ^{-M\beta _2(\omega )}\right. 
\nonumber \\
&&+M[e^{M\beta _1(\omega )}\tau ^{M\beta _2(\omega )-\frac 12}-e^{-M\beta
_1(\omega )}\tau ^{-M\beta _2(\omega )-\frac 12}]h(\omega ;\tau ,r)\left.
{}\right] -\ln (2M) \\
b(\tau ,r) &\sim &\frac 12c_1+M\frac{e^{2M\beta _1(\omega )}\tau ^{2M\beta
_2(\omega )}[1+2M\tau ^{-\frac 12}h(\omega ;\tau ,r)]-1}{e^{2M\beta
_1(\omega )}\tau ^{2M\beta _2(\omega )}[1+2M\tau ^{-\frac 12}h(\omega ;\tau
,r)]+1}
\end{eqnarray}

\section{Duality}

By means of dimensional reduction we can show that the low-energy effective
action (2) is invariant under global $O(d,d)$ transformations, where $d\leq
D $ refers to the number of coordinates it does not depend on [20]. So, if
one assumes a spacetime of the form $N\times K$ where $N$ is a $(D-d)$%
-dimensional spacetime with coordinates $x^\mu $ ($\mu =0,1,..D-d-1$), and $%
K $ a $d$-dimensional compact space with coordinates $y^\alpha $ ($\alpha
=1,..,d)$, and furthermore that all fields are assumed to be independent of
the $y$-coordinates of the ``internal'' space $K,$ then using the notation
of [20] we can rewrite (2) as 
\begin{eqnarray}
S=\int_Ndx\int_Kdy\sqrt{-\hat g}e^{-\hat \phi }(\hat R(\hat g)+\hat g^{\hat
\mu \hat \nu }\partial _{\hat \mu }\hat \phi \partial _{\hat \nu }\hat \phi
-\frac 1{12}\hat H^{\hat \mu \hat \nu \hat \lambda }\hat H_{\hat \mu \hat
\nu \hat \lambda }).
\end{eqnarray}
The hatted quantities now refer to the $D$-dimensional spacetime. Using the
vielbein formalism, $\hat g_{\hat \mu \hat \nu }$ is written as 
\begin{eqnarray}
\hat g_{\hat \mu \hat \nu }=\left( 
\begin{array}{lr}
g_{\mu \nu }+A_\mu ^{(1)\gamma }A_{\nu \gamma }^{(1)} & A_{\mu \beta }^{(1)}
\\ 
A_{\nu \alpha }^{(1)} & G_{\alpha \beta }
\end{array}
\right)
\end{eqnarray}
where $g_{\mu \nu }$ is the metric on $N$ and $G_{\alpha \beta }$ the metric
on $K$.

Define a shifted dilaton 
\begin{eqnarray}
{\hat \Phi }\equiv \hat \phi -\frac 12\log detG_{\alpha \beta },
\end{eqnarray}
and a $2d\times 2d$ matrix $Q$, written in $d\times d$ blocks,

\begin{eqnarray}
Q\equiv \left( 
\begin{array}{lr}
G^{-1} & -G^{-1}B \\ 
BG^{-1} & G-BG^{-1}B
\end{array}
\right) .
\end{eqnarray}
It can be shown that (121) is invariant under global $O(d,d)$
transformations 
\begin{eqnarray}
\hat \Phi \rightarrow \hat \Phi \;\;\;\;\;Q\rightarrow \Omega Q\Omega ^T
\end{eqnarray}
where $\Omega \in O(d,d)$, that is $\Omega ^T\eta \Omega =\eta $ where 
\[
\eta =\left( 
\begin{array}{lr}
0 & {\rm I}\!{\rm I}_d \\ 
{\rm I\!I}_d & 0
\end{array}
\right) 
\]
and ${\rm I\!I}_d$ is the $d$-dimensional unity matrix.

In the case of a diagonal $\hat g_{\hat \mu \hat \nu }$ and a vanishing $B$%
-field, with the choices $d=D-1$ and $\Omega =\eta ,$ the scale-factor
duality is recovered. This was first discussed by Veneziano [21]. In this
case the duality transformation results in an inversion of the scale-factors
in the string frame. For a comprehensive discussion of target-space duality
see ref. [22].

In the case of the Einstein-Rosen metric, (9), considered here the
low-energy effective action (2) is invariant under $O(2,2)$ transformations
for $D=4$. In this section the antisymmetric tensor potential $B_{\mu \nu }$
is assumed to be vanishing. Transforming the metric (9) to the string frame,
the ``internal'' metric $G_{\alpha \beta }$ is found to be 
\begin{eqnarray}
G_{\alpha \beta }=\left( 
\begin{array}{lr}
Re^{2\psi }e^\Phi & 0 \\ 
0 & Re^{-2\psi }e^\Phi
\end{array}
\right)
\end{eqnarray}

The shifted dilaton defined above remains invariant under $O(d,d)$
transformations and this implies that the dilaton itself transforms as 
\begin{eqnarray}
\Phi \rightarrow \Phi -\frac 12\ln \frac{detG_{\alpha \beta }}{detG_{\alpha
\beta }^{dual}}
\end{eqnarray}
where the index ``$dual$'' indicates an $O(d,d)$ transformed quantity.

\subsection{Generalized Scale-Factor Duality}

As mentioned above, picking $\Omega =\eta $ results in the case of a
diagonal metric depending on just one (time-like) coordinate and results in
scale-factor duality. So choosing $\Omega =\eta $ in the Einstein-Rosen
case, where there is dependence on time and space variables, could be called
a generalized scale-factor duality.

Transforming $Q$ according to (125) results in 
\begin{eqnarray}
G\rightarrow G^{-1},
\end{eqnarray}
so that 
\begin{eqnarray}
R^{dual}e^{2\psi ^{dual}}e^{\Phi ^{dual}} &=&R^{-1}e^{-2\psi }e^{-\Phi } \\
R^{dual}e^{-2\psi ^{dual}}e^{\Phi ^{dual}} &=&R^{-1}e^{2\psi }e^{-\Phi },
\end{eqnarray}
which implies 
\begin{eqnarray}
\psi ^{dual} &=&-\psi \\
R^{dual} &=&R^{-1}\exp (-\Phi -\Phi ^{dual}).
\end{eqnarray}

Using (127) to find the transformed dilaton gives 
\begin{eqnarray}
\Phi ^{dual}=-\Phi -2\ln R,
\end{eqnarray}
and hence

\begin{eqnarray}
R^{dual}=R.
\end{eqnarray}

It can be explicitly checked that equations (27), (34) and (35) are
invariant under changes to the dual quantities. Equations (27), (28), (34)
and (35) provide the integrability conditions for equations (36) and (37)
which in turn determine the function $\chi $. Since the integrability
conditions are invariant under the above transformation, the equations
remain integrable and by substituting the dual quantities into (36) and (37) 
$\chi ^{dual}$ is found to be 
\begin{eqnarray}
\chi ^{dual}=\chi -2\psi +\Phi +\ln R+C
\end{eqnarray}
where $C$ is a constant.

The dual metric functions (96)-(98) are found as follows 
\begin{eqnarray}
A_1^{dual} &=&e^CRe^\Phi A_1, \\
A_2^{dual} &=&A_3, \\
A_3^{dual} &=&A_2.
\end{eqnarray}

Since $b(\tau ,r)=0$ in this section, $\Phi (\tau ,r)$ satisfies an equation
similar to that for $\psi $. To find the corresponding Kasner exponents (cf
section 5.1) of the dual model it is necessary to set $M=1$ and $^0\omega
_1=^0\phi _1$ in the equations of section (5.1). This reduces to

\begin{eqnarray}
A_{1}^{dual}&\sim&\tau^{\frac{3}{4}+^{0}\Phi_{1}+\frac{1}{4}
(^{0}\Phi_{1})^{2}+(^{0}\psi_{1})^{2}} \\
A_{2}^{dual}&\sim&\tau^{\frac{1}{2}-^{0}\psi_{1}} \\
A_{3}^{dual}&\sim&\tau^{\frac{1}{2}+^{0}\psi_{1}}.
\end{eqnarray}

Changing to proper time and reading off the Kasner exponents as described in
section (5.1) results in 
\begin{eqnarray}
\sum_{i=1}^{3}p_{i}=1 \\
\sum_{i=1}^{3}p_{i}^{2}=1-2\frac{1+^{0}\Phi_{1}+\frac{1}{4}
(^{0}\Phi_{1})^{2}}{[\frac{7}{4}+\;^{0}\Phi_{1} +\frac{1}{4}%
(^{0}\Phi_{1})^{2}+(^{0}\psi_{1})^{2}]^{2}}.
\end{eqnarray}

This shows that the behaviors of the original and dual model are very
similar. This is expected since only $A_1$ really changes and $A_2$ and $A_3$
are just exchanged.

At this point, it should be mentioned that the usual general-relativistic
constraint on the sum of the Kasner exponents defining the quasi-Kasner
behavior is recovered (cf equations (109) and (142)) since we are working in
the Einstein frame. Assuming a Bianchi I background in the Einstein frame
and transforming the Kasner solutions from the Einstein to the string frame
results in a constraint on the sum of the squares of the Kasner exponents
being unity. This behavior is characteristic for Kasner-like solutions in
the string frame [23]. This, in a way, is more illuminating, since it
reflects directly the invariance under scale factor duality which implies
the (discrete) transformation of a Kasner exponent to its negative ($%
p_i\rightarrow -p_i$).

\section{Discussion}

We have shown that it is possible to find exact inhomogeneous cosmological
solutions of low-energy string cosmology. These solutions are cylindrically
symmetric and represent cylindrical axionic, dilatonic, and gravitational
waves propagating inhomogeneously on a flat anisotropic background. When the
inhomogeneities are of small amplitude these solutions will approach the
behavior of small perturbations of isotropic and homogeneous anisotropic
string models. These solutions also allow us to study the evolution of the
universe in two physically distinct limits: when the inhomogeneities are
larger or smaller than the particle horizon. The behavior found has a simple
physical interpretation. When inhomogeneities are larger than the horizon
they evolve quasi-homogeneously but when they enter the horizon there is
time for self-interaction to occur and the inhomogeneities oscillate like
waves. The axion and dilaton fields behave like two fluids in which the sound
speed equals the speed of light and so shock waves do not form even when the
non-linearities are of large amplitude. The global structure of our
solutions prevents the formation of gravitationally trapped regions and so
there is no primordial black hole formation. (If the $S^3$ topology had been
chosen, with the associated choice $R=\sin \tau ,$ then this would have been
possible).

Solutions of varying degree of generality to (3+1)-dimensional string
cosmology with dilaton and axion in a spacetime of cylindrical symmetry have
been discussed. We found that, in general, the axion-dilaton system is
decoupled from the gravitational background by the cylindrical symmetry.
However, the solutions of section (3.3) are special in a sense that they
describe a universe at large $\tau $ which contains scalar and gravitational
waves that are coupled by the wave-like solutions in the axion field. The
most general Charach-type solutions describe at large values of $\tau $ an
anisotropic universe filled with gravitational and scalar waves caused by
the dynamics of the axion and dilaton. These two regimes also allow us to
find the asymptotic behavior of the universe as $\tau \rightarrow 0$ and $%
\tau \rightarrow \infty .$ There is an initial curvature singularity where
the density of the dilaton and axion fields is formally infinite (hence we
venture outside of the low-energy string theory regime assumed here). The
early-time behavior resembles the Kasner singularity of general relativity
with spatially varying indices and is analogous to that observed on scales
larger than the horizon at later times. The late-time evolution cannot
straightforwardly be compared with the present universe because of the
absence of fermionic fields which provide the standard matter and radiation
components of the Big Bang model. The impact of duality upon these solutions
is more subtle than in the cosmological models that have been examined
previously in string theory because of the presence of inhomogeneity. This
was discussed in detail in section 6 together with the relationships between
the results in the Einstein and string frames.

In summary: we have found exact inhomogeneous and anisotropic cosmological
solutions of low-energy string theory with non-zero axion and dilaton
stresses. These provide a new theoretical laboratory in which to explore the
ramifications of low-energy string cosmology and to use as a basis for
incorporating the effects of higher-order corrections.

\vspace{0.5 cm}

{\bf Acknowledgements }The authors would like to thank E. Copeland, M.
Dabrowski, M. Hindmarsh, and J. Isenberg for discussions. JDB was supported
by a PPARC Senior Fellowship and KEK was supported by the German National
Scholarship Foundation.\\

\section*{Appendix}

Using the properties of the Bessel and Neumann functions [19] the limits for 
$\psi(\tau,r)$ (or $\omega(\tau,r)$) and $G(\tau, r)$ are found.

\subsection*{The limit $n\tau\ll 1$}

\[
\psi (\tau ,r)\sim \alpha _1(\psi ;r)+\alpha _2(\psi ;r)\ln \tau 
\]
where 
\[
\alpha _1(\psi ;r)\equiv \;^0\psi _0+\sum_{n=1}^\infty \cos
[n(r-r_n)][^A\Psi _n+\frac 2\pi \;^B\Psi _n\ln \;n] 
\]
\[
\alpha _2(\psi ;r)\equiv \;^0\psi _1+\frac 2\pi \sum_{n=1}^\infty \cos
[n(r-r_n)]^B\Psi _n 
\]
The function $G(\tau ,r),$ of eq. (47), is found to approach

\[
G(\psi ;\tau ,r)\sim \gamma _1(\psi ;r)+\gamma _2(\psi ;r)\ln \tau 
\]
where 
\[
\gamma _1(\psi ;r)\equiv ^0\psi _0+^0\psi _1\sum_{n=1}^\infty \cos
[n(r-r_n)][^A\Psi _n+^B\Psi _n\ln \;n] 
\]
\[
\gamma _2(\psi ;r)\equiv \frac 12(^0\psi _1)^2+^0\psi _1\sum_{n=1}^\infty
\cos [n(r-r_n)]^B\Psi _n. 
\]

\subsection*{The limit $n\tau\gg 1$}

\[
\psi (\tau ,r)\sim \beta _1(\psi )+\beta _2(\psi )\ln \tau +\tau ^{-\frac
12}h(\psi ;\tau ,r) 
\]

where 
\[
\beta _1(\psi )\equiv \;^0\psi _0 
\]
\[
\beta _2(\psi )\equiv \;^0\psi _1 
\]
\[
h(\psi ;\tau ,r)\equiv \sum_{n=1}^\infty \left( \frac 2{\pi n}\right)
^{\frac 12}\cos [n(r-r_n)][^A\Psi _n\cos (n\tau -\frac \pi 4)+^B\Psi _n\sin
(n\tau -\frac \pi 4)] 
\]

For $G(\psi ;\tau ,r),$ the limiting behavior is found to be

\[
G(\psi ;\tau ,r)\sim \gamma _3(\psi )+\gamma _4(\psi )\ln \tau +\gamma
_5(\psi )\tau 
\]
where 
\[
\gamma _3(\psi )\equiv ^0\psi _0 
\]
\[
\gamma _4(\psi )\equiv \frac 12(^0\psi _1)^2 
\]
\[
\gamma _5(\psi )\equiv \frac 1{2\pi }\sum_{n=1}^\infty n[(^A\Psi
_n)^2+(^B\Psi _n)^2] 
\]

{\bf \ References}\\

[1] M. B. Green, J. H. Schwarz and E. Witten {\it Superstring Theory,}{\sl \ 
}Vol. I (CUP: Cambridge, 1987); E. S. Fradkin, and A. A. Tseytlin, Nucl.
Phys. {\bf B 261,} 1 (1985); C. G. Callan, E. J. Martinec and M. J. Perry,
Nucl. Phys. {\bf B} {\bf 262}, 593 (1985); C. Lovelace, Nucl. Phys.{\bf \ B} 
{\bf 273}, 413 (1985).

[2] E. J. Copeland, A. Lahiri, and D. Wands, Phys. Rev. D {\bf 50}, 4868
(1994).

[3] E. J. Copeland, A. Lahiri, and D. Wands, Phys. Rev. D {\bf 51,} 1569
(1995)

[4] M. Gasperini, J. Maharana and G. Veneziano, Phys. Lett. B {\bf 272}, 277
(1991); M. Gasperini, and G. Veneziano, Phys. Lett. B {\bf 277}, 256 (1992);
M. Gasperini, R. Ricci and G. Veneziano, Phys. Lett. B {\bf 319}, 438
(1993); M. Gasperini and R. Ricci, Class. Quantum Grav. {\bf 12}, 677 (1995).

[5] N.A. Batakis and A.A. Kehagias, Nuc.Phy.B{\bf 449,}2481995); N.A.
Batakis, Phys. Lett. B{\bf 353,}450(1995).

[6] N. A. Batakis, Nucl. Phys. B {\bf 353,} 39 (1995); Class. Q. Grav. {\bf %
13,} L95 (1996).

[7] J.D. Barrow and K.E. Kunze, Phys. Rev. D {\bf 55,} 623-9 (1997).

[8] J.D. Barrow and M. Dabrowski, Phys. Rev. D {\bf 55,} 630-8 (1997).

[9] P.J. Adams, R.W. Hellings, R.L. Zimmerman, H. Farhoosh, D.I. Levine, and
S. Zeldich, Astrophys. J. {\bf 253}, 1 (1982); M. Carmeli, Ch. Charach and
A. Feinstein, Ann. Phys. {\bf 150}, 392 (1983); J.L. Hanquin and J. Demaret,
J. Phys. A {\bf 16}, L5 (1983); Class. Q. Grav. {\bf 1}, 291 (1984).

[10] A. Einstein and N. Rosen, J. Franklin Inst.{\bf \ 223}, 43 (1937); A.S.
Kompanyeets, Sov. Phys. JETP {\bf 7}, 659 (1958).

[11] M. Carmeli, Ch. Charach, and S. Malin, Phys. Rep. {\bf 76,} 79 (1981);
A. Krasi\'nski, {\it Physics in an Inhomogeneous Universe,} (N. Copernicus
Astronomical Center, Warsaw, 1993); E. Verdaguer, Phys. Rep. {\bf 229,} 1
(1993).

[12] K.A. Khan and R. Penrose, Nature {\bf 229, }185 (1971); P. Szekeres, J.
Math. Phys. {\bf 13,} 286 (1972); J.B. Griffiths, {\it Colliding Plane Waves
in General Relativity, }(Oxford UP, Oxford, 1991).

[13] K. Tomita, Prog. Theor. Phys. {\bf 59,} 1150 (1978).

[14] R. Gowdy, Phys. Rev. Lett. {\bf 27}, 827 (1971); Ann. Phys. (NY) {\bf 83%
}, 203 (1974).

[15] P.T. Chrusciel, J. Isenberg and V. Moncrief, Class. Q. Grav. {\bf 7},
1671 (1990).

[16] Ch. Charach, Phys. Rev. D {\bf 19,} 3516 (1979).

[17] Ch. Charach and S. Malin, Phys. Rev. D {\bf 21,} 3284 (1980); M.
Carmeli and Ch. Charach, Phys. Lett. A {\bf 75,} 333 (1980).

[18] E. P. Liang, Astrophys. J., {\bf 204,} 235 (1976).

[19] M. Abramowitz and I.A. Stegun {\it Handbook of Mathematical Functions,}
Dover, NY{\it ,} (1964)

[20] J. Maharana, J.H. Schwarz, Nucl.Phys. B {\bf 390}, 3 (1993)

[21] G. Veneziano, Phys. Lett. B {\bf 265}, 287 (1991); K.A. Meissner and G.
Veneziano, Phys. Lett. B {\bf 267}, 33 (1991); M. Gasperini, J. Maharana and
G. Veneziano, Phys. Lett. B {\bf 272}, 277 (1991)

[22] A. Giveon, M. Porrati and E. Rabinovici, Phys. Rep. {\bf 244}, 77 (1994)

[23] M. Mueller, Nucl.Phys. B {\bf 337}, 37 (1990).

[24] J.D. Barrow, Phys. Rev. Lett. {\bf 46}, 963 (1981); J.D. Barrow, Phys.
Rep. {\bf 85},1 (1982), J.D. Barrow and J. Stein Schabes, Phys. Rev. D {\bf %
32,} 1595 (1985).

[25] J.D. Barrow and M. Dabrowski, preprint (1997)

\end{document}